\documentclass[aps,prb,twocolumn]{revtex4}
\usepackage{graphics}
\begin{document}
\title{Charge and spin order in one-dimensional electron systems with
  long-range Coulomb interactions}
\author{B. Valenzuela}
\affiliation{Instituto de Ciencia de Materiales, CSIC,
Cantoblanco, E-28049 Madrid, Spain}
\author{S. Fratini}
\affiliation{Laboratoire d'Etudes des Propri\'et\'es Electroniques
  des Solides, CNRS, 25 avenue des Martyrs, BP\ 166,
F-38042 Grenoble Cedex 9, France}
\author{D. Baeriswyl}
\affiliation{D\'epartement de Physique, Universit\'e de
  Fribourg, P\'erolles,
CH-1700 Fribourg, Switzerland}
\begin{abstract}
We study a system of electrons interacting through
long--range Coulomb forces on a one--dimensional lattice, by
means of a variational ansatz which is the strong--coupling
counterpart of the Gutzwiller wave function. Our aim is to
describe the quantum analogue of Hubbard's classical
``generalized Wigner crystal''. We first analyse charge
ordering in a system of spinless fermions, with particular
attention to the effects of lattice commensurability.  We
argue that for a general (rational)  number of electrons per
site $n$ there are three regimes, depending on the relative
strength $V$ of the long--range Coulomb interaction (as
compared to the hopping amplitude $t$). For very large $V$
the quantum ground state differs little from Hubbard's
classical solution, for intermediate to large values of $V$
we recover essentially the Wigner crystal of the continuum
model, and for small $V$ the charge modulation amounts to a
small--amplitude charge--density wave. We then include the
spin degrees of freedom and show that in the Wigner crystal
regimes (i.e.\ for large $V$) they are coupled by an
antiferromagnetic  kinetic exchange $J$, which turns out to
be smaller than the energy scale governing the charge
degrees of freedom. Our results shed new light on the
insulating phases of organic quasi--1D compounds where the
long--range part of the interaction is unscreened, and
magnetic and charge orderings coexist at low temperatures.
\end{abstract}
\date{\today}
\maketitle

\section{Introduction}

Compounds that are poor electrical conductors or insulators often exhibit
charge patterns which differ from the homogeneous charge
distribution encountered in ordinary metals. The charge ordering often
occurs at a relatively high temperature and is accompanied by a magnetic
ordering at a lower temperature.
These phenomena are quite generic, as they are observed in a variety of
systems. Important examples are the three-dimensional nickelates, the
layered cuprates at specific commensurate fillings, layered molecular
crystals such as BEDT-TTF radical salts
and quasi--one--dimensional systems such as the
organic TMTTF and DCNQI compounds.
From a theoretical point of view,
the full long-ranged Coulomb potential should in principle be
included when dealing with
{\it any} interacting insulating phase,
due to the ineffectiveness of screening.
Accordingly,
it is not surprising to find wide classes of apparently
different compounds showing similar behaviours in their insulating
phases.
In this respect, the possibility that the two phenomena
of charge and spin ordering share a common origin
is very appealing.
Yet, it is not obvious that the two separate energy scales can be described
within the same physical mechanism.

This raises the question of determining a set of
unambiguous signatures of the
long-range Coulomb interactions, to be compared with the experiments.
One such signature is of course charge ordering itself, as was first
recognised by Wigner in the framework of the three-dimensional electron
gas.\cite{Wigner} In the low-density limit, the kinetic energy of the
electrons becomes negligible
and the charges arrange themselves in the form of a {\it Wigner lattice}
which minimizes the electrostatic energy.
However,  it is sometimes difficult to determine
whether the observed charge modulation is induced by
long--range forces or by an instability of the Fermi surface.
Moreover, even if the main driving mechanism
is given by the Coulomb forces,
the presence of other effects (elastic, magnetic) in real compounds can
strongly modify the predicted ordering pattern.

When applying the idea of Wigner crystallization to electrons in
narrow-band solids,  we are faced with the additional constraint that
the charges must sit on the sites of a discrete lattice (the
{\it host} lattice of atoms, to be distinguished from the
Wigner crystal of electrons).
This has been pointed out by Hubbard,\cite{Hubbard} who considered the
extreme limit of zero bandwidth,
a sensible starting point if the dominant energy
scale is set by the interactions.

In the present work, we propose a variational treatment
which allows to go beyond the classical limit by
including the effects of quantum fluctuations.
Our results indicate that
for small but finite hopping parameters,
the charge and spin degrees of freedom are energetically decoupled:
charge crystallization, governed by the (large) Coulomb energy,
naturally gives rise to antiferromagnetic spin correlations,
characterised by a much smaller energy scale.
The coexistence of charge and spin ordering at low energies can thus
be considered as distinctive of dominant long--range Coulomb
interactions, and should therefore
be ubiquitous in the insulating phases of strongly interacting
systems.

It can be argued that for special fillings, short-range
models with one or few interaction parameters (e.g.\ the Hubbard model for
half--filled bands, or the extended Hubbard model for quarter--filled
bands) can give a correct description of the system.
Although they give the right answer for the ground state
charge configuration for special fillings, the
strong--coupling analysis presented here shows that
these models  can no longer be applied as soon as
the filling deviates from such special values, since in
this case they do not yield the proper classical ground
state (a system  with purely local interactions is metallic
away from half--filling). Therefore, a complete
understanding of the behaviour of the insulating phases
as a function of the density can
only be achieved by taking into account the long range part of the Coulomb
potential.

We study a system of fermions in one dimension, interacting via a
local repulsion $U$ and a long--ranged Coulomb potential $V_m=V/|m|$,
$m$ being the distance between electrons.
The corresponding tight--binding Hamiltonian on an L-site ring is
\begin{equation}
  \label{eq:H-LR}
  \hat{H}= -t\hat{T} +  U\hat{D} + V\hat{W}
\end{equation}
where we have defined the following dimensionless operators,
describing respectively electron hopping, on--site and long--range repulsion,
\begin{eqnarray}
 \hat{T} & = & \sum_{n\sigma, s=\pm 1}
c_{n\sigma}^+ c_{n+s \sigma}\;\;,
         \label{kin-op} \\
   \hat{D}&=& \sum_i  n_{i\uparrow}n_{i\downarrow} \;\;,
         \label{double-op} \\
   \hat{W}&=& \frac{1}{2}\sum_{i\neq j \sigma \sigma^\prime} \;
\frac{n_{i,\sigma}n_{j,\sigma^\prime}}{|i-j|} \;\;,
       \label{pot-op}
\end{eqnarray}
$c_{i,\sigma}^+$ ($c_{i,\sigma}$) creates (destroys) an electron
of spin $\sigma$ at site $i$, $n_{i,\sigma}=c_{i,\sigma}^+ c_{i,\sigma}$
is the occupation number of this state and $t$ is the hopping
 amplitude between nearest--neighbor sites.

The paper is organized as follows.
In the next section we introduce our variational wave function, starting
from Hubbard's solution for the classical limit.
In Section \ref{spinless}, we study the
phenomenon of charge ordering at different commensurate fillings
by neglecting the magnetic degrees of freedom.
These are analysed for a quarter--filled band in Section \ref{magnetic},
where contact is also made with experimental results for
organic quasi--one--dimensional compounds.

\section{Variational wave function for generalized Wigner lattices}
\label{sec:GWL}

The limit studied by Hubbard\cite{Hubbard} is obtained by setting
the bandwidth to zero and $U \to \infty$.
In this case the problem is classical (but
still non--trivial), and corresponds to minimising the interaction
energy by distributing the $N$ electrons as homogeneously as possible
over the $L$ available lattice sites.
The difficulty arises from the competition between the different
interaction parameters $V_m$. If the potential is convex, as in the
Coulomb case, a solution can be found for any commensurate filling
$n=N/L=r/s$ ($r$ and $s$ integers) in the form of the
generalized Wigner lattice (GWL), a periodic sequence of unit cells
of size $s$, each containing $r$ electrons distributed according to a
minimum spread of interelectron distances.
The charge density on site $l$ is either $1$ or $0$
according to the following formula:\cite{Aubry}
\begin{equation}
  \label{eq:aubry}
  n_l= [l r/s] - [(l-1)r/s]
\end{equation}
where $[\ldots]$ gives the integer part of the argument.

For finite $t$ quantum fluctuations will tend to delocalize the electrons away
from the classical configuration.
Nevertheless, the classical solution is still a good approximation for
small $t$ where most of the charge is located
on the sites predicted by eq.\ (\ref{eq:aubry}). This argument can be
made more precise for simple fillings $n=1/s$.
Moving a particle away from the classical configuration costs an energy
\begin{equation}
  \label{eq:dipolar}
  \Delta =2 V \sum_{m=1}^\infty \frac{1}{ms[(ms)^2-1]}\ .
\end{equation}
A density $\delta n\sim (t/\Delta)^2$ will thus be allowed to leak to the
nearest unoccupied sites.
The charge gap is $\Delta =0.39V$ at quarter filling ($n=1/2$) and is
rapidly reduced at lower fillings,
$\Delta  \simeq  2.4 n^3 V$, since the repulsion that prevents
the occupation of ``wrong'' sites weakens as the inter--electron
distance increases.
The range of validity of the classical approximation can be estimated
by requiring $\delta n$ to be smaller than a threshold value $P$.
This leads to the phenomenological criterion that
quantum band motion has negligible effects as long as $V$ exceeds
the value
\begin{equation}
  \label{eq:Vcl}
  V_{GWL}\approx \frac{t}{2.4 \sqrt{P} n^3}\ .
\end{equation}
Note that $V_{GWL}\sim n^{-3}$, so that
it is harder to stabilize a classical GWL at low densities,
where $V_{GWL}$ can be very large.
Taking for instance $P=0.2$, eq. (\ref{eq:Vcl}) gives $V_{GWL}/t\approx
8$ at $n=1/2$ and $V_{GWL}/t\approx 500$ at $n=1/8$.
\footnote{For more complex fillings such as $n=r/s$, the expression
  for the gap is not as simple as eq. (\ref{eq:dipolar}). The
occupied sites are not all equivalent, and leakage to a wrong site now
depends crucially on the local environment. Expression
(\ref{eq:Vcl}) can still be considered as a rough estimate for the
general case, although sensible
variations can be expected when moving from simple
commensurabilities such as $n=1/2$ or $n=1/3$ to more complex charge patterns.}

In order to study the ground state of the full Hamiltonian
(\ref{eq:H-LR}) we now use the variational ansatz
\begin{equation}
  \label{eq:Psi_B}
  |\Psi_B(\eta)\rangle= e^{-\eta \hat{T}}|\Psi_\infty\rangle\ ,
\end{equation}
introduced previously as
the strong coupling counterpart of the Gutzwiller wave function in the
context of the Mott--Hubbard transition\cite{Baer}. Here
$\eta$ is a dimensionless variational parameter,
$\hat{T}$ is the kinetic energy operator of eq.\ (\ref{kin-op})
and $|\Psi_\infty\rangle$ is the ground state for $t=0$,
i.e.\ Hubbard's classical solution.
In the same way as the
Gutzwiller wave function reduces double occupancy and thus states with
high potential energy, the factor
$e^{-\eta \hat{T}}$ suppresses configurations with high kinetic energy.

It can be demonstrated that the wave function (\ref{eq:Psi_B})
is charge ordered and insulating for any finite  $\eta$  --- its
charge stiffness, or equivalently, its Drude weight, vanishes.\cite{Dzi}
A transition to a metallic state occurs for $\eta\rightarrow \infty$,
where only the configuration with the lowest kinetic energy, i.e.\ the filled
Fermi sea, survives.
$|\Psi_B\rangle$ is expected to provide a fairly
accurate description of charge fluctuations in the strong coupling limit,
but it can no longer be trusted for weak couplings, although it tends to
the correct limit as the interaction vanishes.
\footnote{For weak interactions, a better estimate for the
ground--state energy can in principle be obtained using
$$  |\Psi_G(\lambda)\rangle= e^{-\lambda \hat{W}}|\Psi_0\rangle\ , $$
where $|\Psi_0\rangle $ is the Fermi sea.
The operator $e^{-\lambda \hat{W}}$ builds up the
charge correlations appropriate for the GWL, by enhancing the
configurations with low interaction energy. This wave function can be
considered as  the ``dual'' of $|\Psi_B\rangle$ and is always metallic.
Direct comparison of the two complementary approaches is expected to give a
good description of charge ordering. However, the
calculation of the variational energy $E_G(\lambda)$ is not simple
when dealing with long--range interactions, and will be left for future
work.}

Our wave function is easy to handle for $U\rightarrow\infty$ where
spin fluctuations are suppressed and the problem becomes equivalent to that
of spinless fermions. In this case, our variational procedure can be used
straightforwardly to study not only simple fillings such as
$n=1/2$ but also much more complex GWL ground states such as
$n=11/47$,  and it can readily  be generalized to higher dimensions.
For finite values of $U$, where spin fluctuations have to be taken into
account,
the problem becomes more complicated. However, in the strong coupling limit
$U,V\gg t$ the energy scales for charge and spin degrees of freedom are
very different, and the problem of magnetic ordering can be addressed
in terms of a low--energy spin Hamiltonian, in fact the simple
antiferromagnetic
Heisenberg model for the case $n=1/2$. This will be discussed in section
\ref{magnetic}.

\section{Spinless fermions}
\label{spinless}
We shall first assume that the on--site repulsion is the dominant
energy scale, and take the limit $U\to\infty$. This eliminates both double
occupancy and mixing of different spin configurations. We can therefore
suppress the spin index in the Hamiltonian
(\ref{eq:H-LR})-(\ref{pot-op}), and ignore the local term (\ref{double-op}).
Transforming to Fourier space leads to the following simplified model
for
{\it spinless} electrons:
\begin{equation}
\hat{H}= -t\hat{T}+ V\hat{W} =
\sum_k \epsilon_k c_k^+ c_k + \frac{1}{2L}\sum_{q} {}^\prime
V(q) \rho_{q}\rho_{-q}
\label{eq:H-LR-spinless}
\end{equation}
where the band dispersion relation is $ \epsilon_k=-2t\cos k$,
$\rho_{q}=\sum_{k} c^+_{k+q } c_{k}$
is the density fluctuation
operator, and  $V(q)=-V \log [2(1-\cos q)]$ (the lattice
parameter has been set equal to 1).
The $q$ sums run over the entire Brillouin zone except $q=0$, which
ensures charge neutrality. To avoid multiple counting of the
interaction energy,
we have taken the limit $L\rightarrow \infty$
with the prescription that the range of the interaction
potential does not
exceed half the length of the ring.
\footnote{The distance $m$ between sites $i$ and $j$  is defined as
$m=|i-j|$, the linear distance.
Other choices are possible. For instance, taking $m$ to be the chord
distance on the circle [G. Fano, F. Ortolani, A. Parola and L. Ziosi,
  Phys. Rev. B {\bf 60}, 15654 (1999); Y. Hatsugai,
Phys. Rev. B {\bf 56}, 12183 (1997)]
slightly modifies the function $V(q)$ --- which can no longer be expressed in
closed analytical form --- but has the advantage of
avoiding multiple counting by construction. }
The Coulomb potential has a
logarithmic singularity at long wavelengths, characteristic of the
one-dimensional case, and an intrinsic cut-off
at short wavelengths given by the lattice parameter.

The wave function (\ref{eq:Psi_B}) can be worked out exactly for any
filling $n=r/s$. We shall present in full detail the formalism for
$n=1/2$. Results will be given for different fillings as well.

\subsection{Filling $n=1/2$}
\label{filling-quarter}
The classical solution in this case corresponds to alternating occupied
and empty sites. This can be written in k--space as:
\begin{equation}
  |\Psi_\infty\rangle=
\prod_{k\in RBZ}\frac{1}{\sqrt 2} (c^+_k +  c^+_{k+\pi}) |0\rangle
\end{equation}
where the product runs over the reduced
Brillouin zone $|k|<\pi/2$, and $|0\rangle$ is the
vacuum for electrons. The phase correlator
$e^{-\eta \hat{T}}$ is now diagonal and we can write straightforwardly
the (normalized) variational wave function (\ref{eq:Psi_B}) as
 \begin{equation}
  |\Psi_B(\eta)\rangle =
\prod_{k \in RBZ}\frac{1}{N_k}
(e^{-\eta \epsilon_k}c^+_k +  e^{\eta \epsilon_k}c^+_{k+\pi}) |0\rangle
\end{equation}
where the normalization factor is given by  $N_k^2=2 \cosh (2 \eta
\epsilon_k)$ (we have set $t=1$ so that energies are now expressed
in units of the hopping parameter).
The best variational ground state is obtained by minimising the energy
functional
\begin{equation}
  E_B(\eta)  =  \langle\Psi_B(\eta)| \hat{H} |\Psi_B (\eta)\rangle
\label{e-b-sl}
\end{equation}
with respect to the variational parameter.
The kinetic energy per unit
length is readily evaluated
\begin{equation}
  \label{eq:kinetic}
  \langle\hat{T}\rangle/L =\int_{RBZ} \frac{dk}{2\pi} \; \epsilon_k \tanh
(2\eta \epsilon_k).
\end{equation}
The potential energy can be expressed in terms of the
structure factor $S(q)=\langle\rho_q\rho_{-q}\rangle/L$,
\begin{equation}
  \label{eq:potential}
   \langle V\hat{W}\rangle/L=\frac{1}{4\pi}\int  dq S(q) V(q).
\end{equation}
The structure factor has a regular part
\begin{equation}
  \label{eq:struct-reg}
  S(q)=\frac{1}{4}-\frac{1}{4\pi} \int_{RBZ} dk
\frac{1+\sinh (2\eta \epsilon_k) \sinh (2\eta  \epsilon_{k-q})}
{\cosh  (2\eta \epsilon_k ) \cosh (2\eta \epsilon_{k-q})}
\end{equation}
and a Bragg peak at $q=\pi$ corresponding to the periodic ordering of the
charges:
\begin{equation}
  \label{eq:struct-Bragg}
  S(\pi)=L\left\lbrack \int_{RBZ} \frac{dk}{2\pi} \frac{1}{\cosh  (2\eta
\epsilon_k
      )}\right\rbrack^2.
\end{equation}
Writing the average density distribution
as $n(l)=n + \tilde{n} \cos (\pi l)$, we see that
the intensity of the Bragg peaks in the
structure factor is proportional to the square of the order parameter,
$S(\pi)/L=\tilde{n}^2$.
It is finite for any finite $\eta$, so that
the wave function $|\Psi_B\rangle$ always represents a charge--ordered state.

\begin{figure}[htbp]
\resizebox{7cm}{!}{\includegraphics{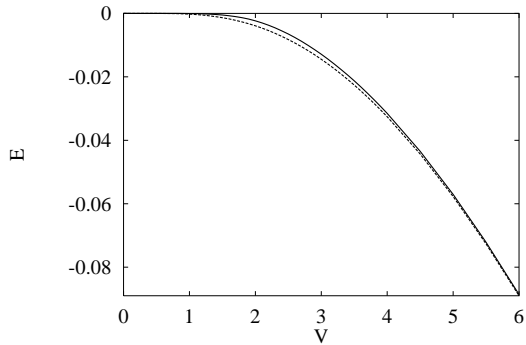}}
\caption[]{Total condensation energy (per unit volume)
  for spinless fermions at $n=1/2$ as a
  function of the interaction strength $V$ (energies are in units of $t$).}
\label{fig:quarter-energy}
\end{figure}

In the limit $\eta\to 0$ one recovers
the classical GWL. The kinetic energy vanishes and
the structure factor becomes $S(q)=(1/4)\delta_{q,\pi}$,
so that equation (\ref{e-b-sl})  gives
$E_{cl}=V(\pi)/8=-(V/4)\log 2$. The opposite
limit $\eta\to\infty$ yields the Hartree--Fock energy of the
liquid phase, which is equivalent to treating the interaction to lowest
order in
perturbation theory. Equation (\ref{eq:struct-reg}) then
gives $S(q)=|q|/2\pi$ and one readily evaluates
$E_{0}(V)=-2/\pi-c V$, with $c=7\zeta(3)/8 \pi^2=0.10657$ ($\zeta$
is the Riemann function).
For finite $\eta$, one can define the condensation energy
as the energy gained through charge ordering,
$E_{cond}=E_B-E_{0}$. This quantity is shown in figure
\ref{fig:quarter-energy} as a function of $V$.
The variational result (full line) is very close to
the Hartree-Fock result in the broken-symmetry phase (dashed line).
The latter is known to be a good approximation in the strong coupling
regime, and is surprisingly accurate in the presence of lattice
commensurability, which strongly reduces charge fluctuations.
\cite{imada02}
Although no phase transition occurs within our variational approach,
there is a crossover between a weak--coupling regime, where the
density modulation is small as compared to the average density, and a
strong--coupling (or classical) regime, where the order parameter approaches
the limiting value of Hubbard's GWL pattern. Correspondingly, the
condensation energy is exponentially small at low $V$, behaving like
$E_{cond}\simeq -0.02 V \exp(-2\pi/V)$, while
at large $V$ it can be expanded in powers of $t/V$ as
$E_{cond}\approx 0.64t-0.067 V -2.59 t^2/V+\ldots$.

\begin{figure}[htbp]
\resizebox{8cm}{!}{\includegraphics{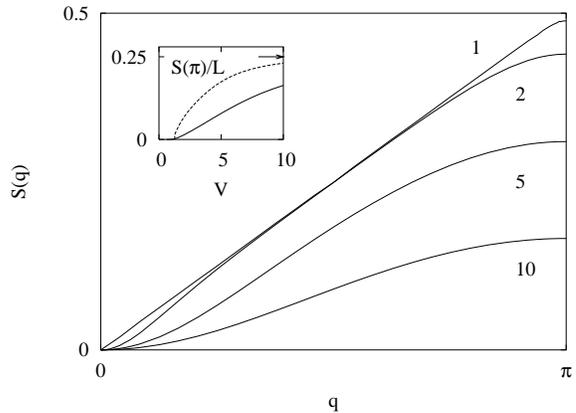}}
\caption[]{Structure factor in the spinless case for different values of
the interaction
  strength at $n=1/2$: from top to bottom, $V=1,2,5,10 t$. The inset shows
the intensity
  of the Bragg peak at $\pi$, proportional to the square of the order
  parameter, versus $V$. The dashed line is the same quantity
  evaluated in a model with only   nearest neighbour interactions. The
  arrow marks the classical limiting value.}
\label{fig:quarter-struct}
\end{figure}
A direct indication on how the charge modulation builds up is given
by the evolution of the structure factor, shown in figure
\ref{fig:quarter-struct}.
At $V=1$ (top line)  $S(q)$ is very close to the structure factor of
non--interacting electrons. Upon increasing $V$, the regular part
is progressively reduced, and some weight is transferred to the Bragg
peak (see the inset, full line).  Despite a marked onset around
$V\approx t$, $S(\pi)$ is still far from its classical limit
(indicated by an arrow) at values as high as $V=10t$.
The variational result for a  model with only nearest-neighbour
interactions (dashed line) shows that for equal values of $V$ in the
charge--ordered regime,
the long--range nature of the Coulomb potential
{\it reduces} the amplitude of charge ordering, because
the energy cost for creating a local charge fluctuation is partly
compensated by the interaction parameters $V_m$ at distances $m>1$.
\footnote{\label{gaps}
The charge gap in the strong--coupling limit can be evaluated
in general for models with $m$--neighbor interactions. The largest value
is obtained for $m=1$ (only n n.\\ interactions), $\Delta_1=V$, and the
lowest is for $m=2$ (n.n.n.\ interactions), $\Delta_2=0$. The gap
for larger $m$ converges to the value $\Delta_\infty=0.39$ given by
eq.\ (\ref{eq:dipolar}) with an oscillatory behavior.}
This is in agreement with refs. \cite{Capponi}, where it is shown that the
metallic character of the system is enhanced when going from short
(i.e.\ nearest--neighbor) to long--range interactions.

\subsection{General rational fillings}

For generic commensurate fillings of the form $n=r/s$, the $s$ harmonics
needed to reproduce the GWL pattern produce $s$ Bragg peaks in the Brillouin
zone at multiples of $2\pi/s$ (for even $s$, the points $\pm \pi$ of the
Brillouin zone are equivalent). It is straightforward to extend the
analysis presented above for $r=1$, $s=2$ to larger values of $s$.
\begin{figure}[htbp]
\resizebox{8cm}{!}{\includegraphics{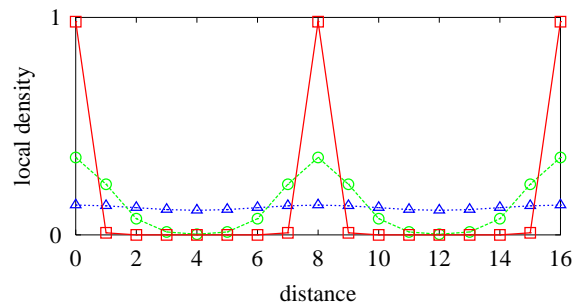}}
\caption[]{Average occupation of site $l$ 
 for different values of the interaction strength at
  $n=1/8$: $V=0.5t$ (triangles), $V=20t$ (circles), $V=500t$
  (squares). Lines are guides to the eye.}
\label{fig:dens}
\end{figure}

Figure \ref{fig:dens} shows the  variational results for the average
site occupation at a filling $n=1/8$, for different values of the interaction
strength. Unlike the case $n=1/2$, which exhibits a single
crossover from weak to strong coupling, three different  regimes can now
be identified. (I) For $V/t=0.5$ (triangles) the modulation of the density
is weak. (II) For larger values of the coupling strength
($V/t=20$,
circles), the charge modulation is strong, and the total wave function is
essentially made up of localised one--electron wave functions with
virtually no overlap, like in an ordinary Wigner crystal. 
The GWL (III), where 
only the sites predicted by Hubbard's solution are occupied, is
achieved only at extremely large $V$ ($V/t=2000$, squares).

\begin{figure}[htbp]
\resizebox{8cm}{!}{\includegraphics{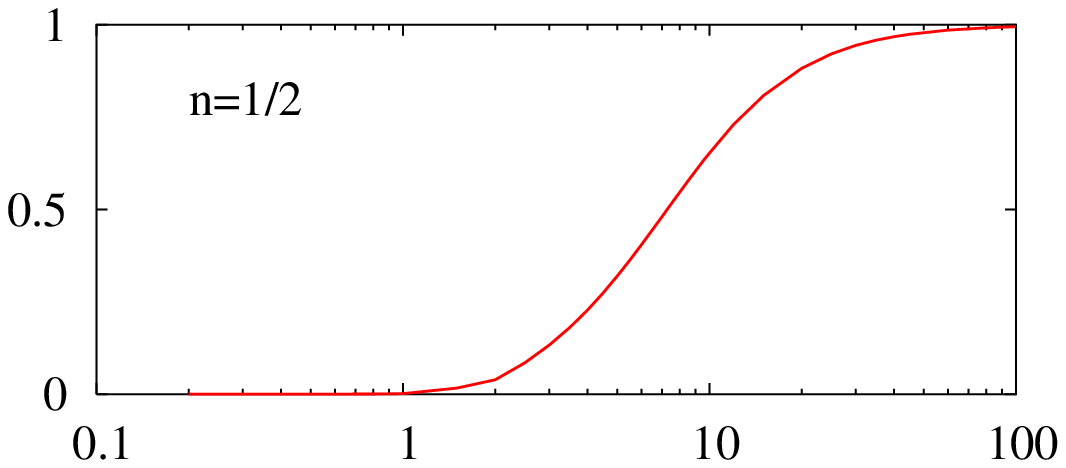}}
\resizebox{8cm}{!}{\includegraphics{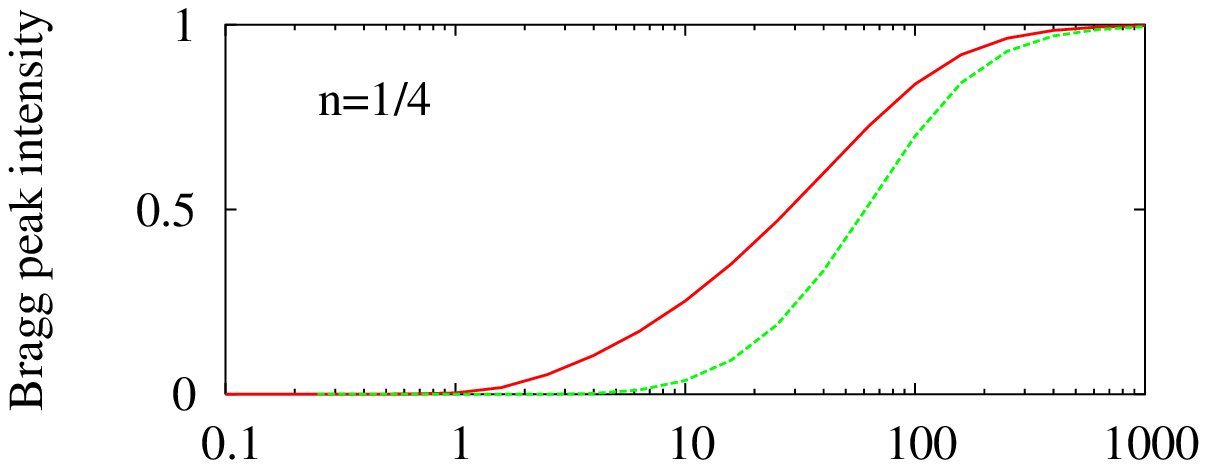}}
\resizebox{8cm}{!}{\includegraphics{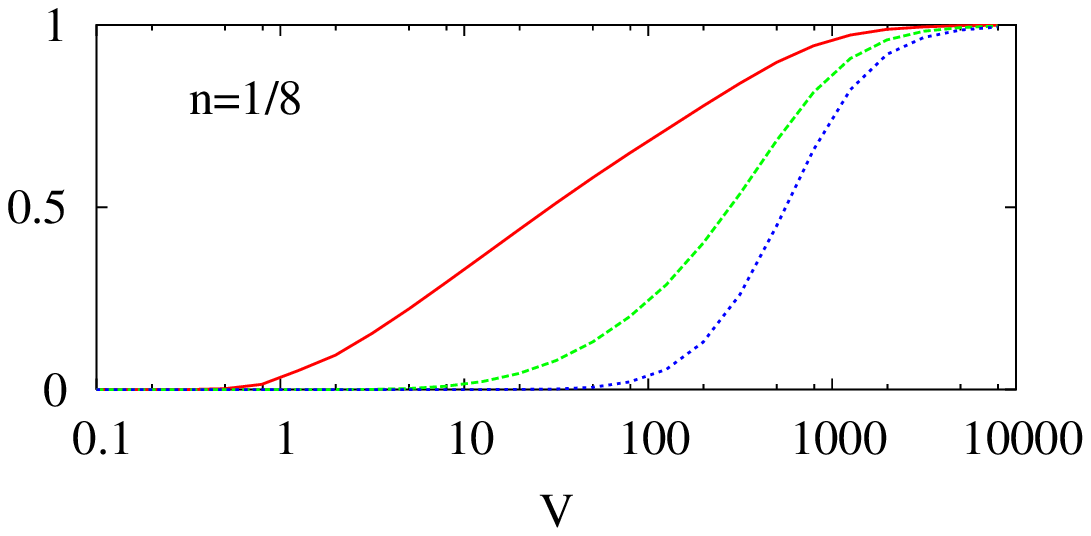}}
\caption[]{Normalised intensities of
  the Bragg peaks at multiples of $q_{WC}=2\pi n$,
  versus the interaction strength. From top to bottom: $n=1/2,1/4,1/8$.  The full, dashed and dotted lines correspond respectively to first, second and third harmonics. Note the different scale in the three graphs.}
\label{fig:bragg}
\end{figure}
Figure \ref{fig:bragg} shows
the normalised intensities of the peaks in the structure factor at
multiples of $q_{WC}=2\pi n$, namely  $S(\nu q_{WC})/n^2 L$, as a
function of the coupling strength, for $n=1/2$ (top), $1/4$ (center)
and $1/8$ (bottom). We
see that at fillings $n<1/2$, higher harmonics
progressively  appear in the structure factor when increasing the
coupling, in addition to the
main peak at $q_{WC}$.  The system goes progressively from 
the weak coupling limit
(I), where the
modulation can be described in terms of a single, exponentially small,
Bragg peak, to  the GWL (III), where all the
harmonics in the structure factor saturate to their classical value.
Note that the intermediate region (II) becomes broader as $n$ decreases. 

\subsection{From the Wigner lattice to the pinned charge--density wave}

By construction, our variational wave function exhibits
long--range order for any finite positive value of $\eta$
(and thus for any positive value of $V$ in the case of the
long--range Coulomb interaction).  The evolution of the charge modulation
proceeds smoothly from a $GWL$ at very large $V$ over an ordinary Wigner
crystal
at large but not too large $V$ to a small--amplitude charge--density wave
from intermediate to small values of $V$. However, it is obviously not clear
whether a trial wave function which is linked to the exact ground state
for $t/V = 0$ can still be trusted in a region where $V/t < 1$, since
in this case it would be more appropriate to start from the filled Fermi
sea.

To examine this point, we consider the soluble model of 1D spinless fermions
with nearest--neighbor interaction (parameter $V$).
For an average occupation $n=1/2$ this model is equivalent to the
$XXZ$ Heisenberg chain in zero magnetic field with exchange constants
$J_x/J_z\ =\ J_y/J_z\ =\ 2t/V$. For $V>2t$ (uniaxial regime for the spin chain)
there is true long--range order, but not for $V<2t$ (planar regime).
The absence of long--range order for small values of $V$ can
also be understood within the Fermi gas model, where the (crystal) lattice
generates an Umklapp term; the latter has to exceed a certain critical
strength in order to sustain long--range charge order \cite{reviewschulz}.
Our wave function predicts long--range order for $V>V_c\approx 1.3t$ and
a weak first--order transition to a metallic phase for $V=V_c$
(see the inset of figure \ref{fig:quarter-struct}). Thus the predictions of
our variational ansatz agree qualitatively with the exact phase diagram.

The question of the existence of true long--range order for 1D lattice
fermions interacting through the $1/r$ Coulomb potential, i.e., for the
Hamiltonian (\ref{eq:H-LR-spinless}), is more subtle and still not completely
understood. In the absence of a lattice potential, Schulz \cite{Schulz}
has shown, using bosonization for small $V$ and an analysis of quantum
fluctuations about the classical Wigner crystal at large $V$, that the
long--distance behavior of the density--density correlation function is
universally given by
\begin{equation}
\langle\rho(x)\rho(0)\rangle \sim
A\ \mbox{cos} (Qx) e^{-\mbox{{\it c}$\sqrt{\log(x)}$}}\ ,
\label{eq:corr}
\end{equation}
where $Q=2k_F$ for the spinless model ($4k_F$ for model (\ref{eq:H-LR}) with
$U=\infty$) and the constant $c$ is independent of $V$. Thus Wigner
crystallization in the sense of quasi--long--range
order exists independently of the interaction strength in one dimension.
This is in sharp contrast to three and two space dimensions, where
Wigner crystallization is known to take place at
{\it finite} values of the interaction strength \cite{Ceperley},
namely at $r_s\simeq 100$
and $r_s\simeq 37$, respectively ($r_s$ is
defined as the radius of a sphere containing one electron in average,
expressed in units of the Bohr radius, and controls the
potential/kinetic energy ratio).

One expects that any additional commensurate lattice potential will
transform the quasi--long--range order into a true long--range order.
This has been confirmed on the basis of exact diagonalization \cite{Capponi}
for large $V$ and $n=1/3$. For small $V$, estimates of the effects of
Umklapp scattering (using bosonization \cite{Capponi}) are also in
line with the existence of a finite order parameter, although its value and
the corresponding lattice pinning energy are expected to be extremely small.

In contrast to bosonization, which cannot provide the amplitude of the
density--density correlation function [the strongly interaction--dependent
constant $A$ in Eq.\ (\ref{eq:corr})], our variational approach allows
us to estimate the order parameter as a function of both the interaction
strength $V$ and the average site occupation $n$. In 1D the dimensionless
parameter of the continuum theory is $r_s=(a/2n)/a_B$.
Within the tight--binding model (\ref{eq:H-LR-spinless}),
the dispersion $\epsilon_k$  close to the band edges
can be approximated by a parabola
with an effective mass $m^*=\hbar^2/(2ta^2)$.
 Using the unscreened Coulomb potential for $V$ and inserting the band
mass $m^*$ into the definition of the Bohr radius, i.e., $a_B = 2ta/V$,
one obtains
\begin{equation}
  \label{eq:rs}
  r_s=\frac{V}{4 n t}\ .
\end{equation}
Starting from the Wigner crystal phase (II), where the
electron wave functions are well separated (cf.\ fig.\
\ref{fig:dens}), and varying $V$, we expect to encounter two
crossovers, the first to a weakly pinned charge--density wave upon
decreasing $V$, the second to a GWL upon
increasing $V$.  The
first crossover towards a weakly modulated ground state takes place
when the spread $\sigma$ of the single--particle wavefunctions equals
a given fraction $\delta$ of the average inter--particle distance $d=1/n$,
namely
\begin{equation}
  \label{eq:Lindemann}
  \frac{\sigma}{d}\approx \delta\ .
\end{equation}
In our lattice problem, the spread of the single--particle
wave functions can be defined as
\begin{equation}
  \label{eq:spread}
  \sigma^2=\sum_{l=-d/2}^{d/2} l^2 n_l
\end{equation}
(the origin has been chosen to coincide with an occupied site of the GWL).
The Lindemann criterion (\ref{eq:Lindemann}) is expected to locate
a broad crossover region rather than a true phase transition in the present
1D case. Therefore the choice of $\delta$ is somehow arbitrary,
and we shall take $\delta=0.25$.

We have calculated the crossover $V_{Lind}$ predicted by the Lindemann
criterion
(\ref{eq:Lindemann}) for various fillings $n=1/32,1/16,1/8,1/4$ and
$1/2$. Remarkably, the values obtained  at low fillings ($n < 1/4$)
scale linearly with $n$, as expected from continuum theory in view
of eq.\ (\ref{eq:rs}), corresponding to a crossover around $r_s\approx 2.5$.
The situation changes at fillings $n=1/4$ and $n=1/2$, where the increasing
influence of lattice effects favors localization of the particles, thus
reducing the value of $r_s$ at the crossover ($r_s\approx 2$ at
$n=1/2$, see also ref.\cite{imada02}). As to the crossover towards the GWL,
we may use the criterion (\ref{eq:Vcl}), i.e.\ $V_{GWL}\sim n^{-3}$.

\begin{figure}[htbp]
\resizebox{8cm}{!}{\includegraphics{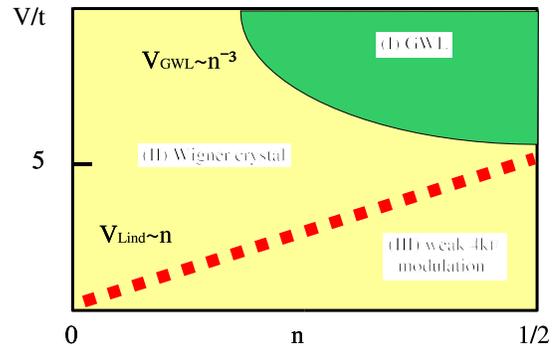}}
\caption[]{ Phase diagram for the spinless model
  (\ref{eq:H-LR-spinless}). A charge--ordered state persists for all
  commensurate fillings with three characteristic regimes, from top to bottom:
  (I) generalized Wigner lattice,
  (II) Wigner crystal and (III) small--amplitude charge--density wave.}
\label{fig:phasediag}
\end{figure}

The  resulting phase diagram is shown in figure \ref{fig:phasediag},
where the two boundaries represent the two crossover regions for
rational fillings $n=1/s,\ s$ integer.
Starting from Hubbard's classical GWL at
$V\to \infty$ (top), the inclusion of quantum fluctuations
at a given density $n$ gives rise to two
successive crossovers. First, around $V_{GWL}$ the particles
begin to spill over to the neighboring lattice sites, and the ground
state becomes similar to the Wigner crystal of the continuum model.
Reducing  further the interaction strength progressively suppresses
the higher harmonics of the density modulation, so that below the
second crossover around $V_{Lind}$ only a small--amplitude ``$4k_F$''
charge--density wave remains.

These crossover lines have opposite density dependences.
$V_{GWL}\sim n^{-3}$ decreases with increasing density, since
reducing the inter--particle distance makes it easier to confine
particles on a single lattice site. $V_{Lind}\sim n$ increases with
density, since the potential/kinetic energy ratio
scales as $V/(nt)$ [eq.\ (\ref{eq:rs})]. At $n=1/2$, where lattice
effects are most important, $V_{GWL}\approx V_{Lind}$ and the two
crossovers become indistinguishable: the Wigner crystal phase is
wiped out, and the system passes directly from the classical GWL to
the weakly modulated phase.
Note that in the spinless case presented here, there is full
particle--hole symmetry relative to $n=1/2$, which allows to deduce
the phase diagram for $1/2<n<1$.

\subsection{Fluctuations of polarization}

\begin{figure}[htbp]
\resizebox{8cm}{!}{\includegraphics{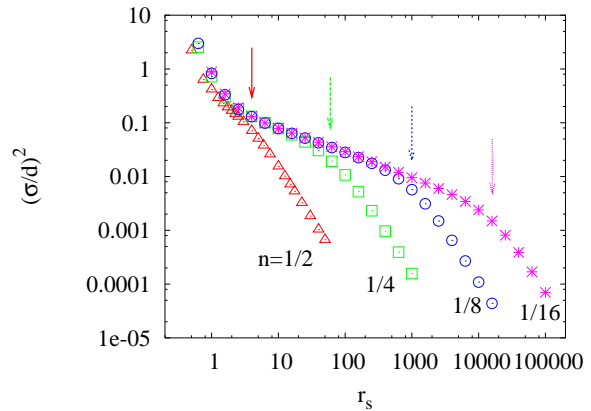}}
\caption{The fluctuations of polarization as a function of  $r_s=V/4nt$,
  for different fillings: $n=1/16$ (asterisks), $n=1/8$ (circles), $n=1/4$
(squares)
and $n=1/2$ (triangles). Scaling breaks down when entering the GWL
phase (marked by arrows), which is dominated by lattice effects.}
\label{fig:polfluc}
\end{figure}

An insulator, such as our electronic system with a
commensurate charge--ordered ground state, responds to an applied electric
field $E$ by a macroscopic polarization $P=\langle \hat{X}/L\rangle$, where
\begin{equation}
  \label{eq:pol}
  \hat{X}=\sum_l l n_l
\end{equation}
is the dipole operator and we have set the electron charge equal to unity.
The dielectric susceptibility is defined as
$\chi=P/E$ in the limit $E\to 0$. The upper bound \cite{Aebischer}
\begin{equation}
  \label{eq:upperbound}
  \chi\le \frac{2}{L\Delta} \langle \hat{X}^2 \rangle
\end{equation}
links $\chi$ to the mean--square fluctuation of polarization
$\langle \hat{X}^2 \rangle$ (we have taken
$\langle \hat{X}\rangle=0$ by symmetry) and to the charge gap
$\Delta$ (defined here as the minimum excitation energy for
which the dipole matrix element does not vanish). In a Wigner solid
the electrons are located close to their  classical equilibrium positions
and each one contributes an amount $\sigma^2$ [cf.\ eq.\ (\ref{eq:spread})]
to the mean--square deviation of polarization, giving
\begin{equation}
  \label{eq:central}
  \frac{\langle\hat{X}^2\rangle}{L}\approx n \sigma^2
\end{equation}
According to the discussion of the previous section,
the properties of the system in region II (the region of the
conventional Wigner crystal) should agree with the predictions of
continuum theory. Therefore, when plotted as a function of $r_s$,
the quantity $(\sigma/d)^2=n\langle\hat{X}^2\rangle/L$ evaluated at
different fillings should fall on a
{\it universal curve}. The results at  $n=1/16$ (asterisks),
$n=1/8$ (circles), $n=1/4$ (squares)
and $n=1/2$ (triangles) are plotted in fig.\ \ref{fig:polfluc}, showing
indeed a scaling behavior up to the value $V_{GWL}(n)$, indicated by
arrows. On the left side of the figure,
the crossover between the Wigner crystal phase (II) and the weakly
modulated regime (I) is signaled by a marked upturn of the curve around
$r_s\approx 2$, which is roughly filling independent (except for $n=1/2$),
and agrees with the phenomenological Lindemann criterion
(\ref{eq:Lindemann}).

The scaling behavior breaks down when we enter the GWL region (III), i.e.\ at
$V_{GWL}$. The onset depends on the filling, so that
the  scaling region is progressively reduced as $n$ increases, and it
completely disappears at $n=1/2$, where
lattice effects are most important. Note that at this particular
filling the fluctuations of polarization are systematically reduced
with respect to the universal curve (the reduction is approximately
$30\%$ around $r_s\approx 2$), showing that lattice
commensurability favors charge localization.
Above $V_{GWL}$ (on the right side of the figure), where the charges
become localized on individual sites,
there is a rapid drop in the fluctuations of
polarization, and it can be shown following the perturbative
arguments of section
\ref{sec:GWL} that $(\sigma/d)^2\simeq 1/(46\ r_s^2 n^6)$.

\section{Magnetic correlations}
\label{magnetic}
The assumption of infinite $U$ implies for the present one--dimensional
case that the different spin configurations do not mix and that the
magnetic exchange constant is zero. It follows that from a thermodynamic
point of view the spins behave as free magnetic moments giving rise to
a Curie--type magnetic susceptibility. In some of the organic materials
such a behavior is indeed observed at temperatures where charge ordering
sets in, but at lower temperatures antiferromagnetic correlations lead to a
substantial reduction of the magnetic susceptibility and finally to a
magnetically ordered ground state. In this section we show that our
variational approach can be readily generalized to finite values of $U$.
We limit ourselves to large Coulomb interactions, where the magnetic energy
scale turns out to be much smaller than the energies associated with
the charge ordering, and treat explicitly the quarter--filled case
($n=1/2$), which is relevant for some
quasi--one--dimensional compounds such as the TMTTF
salts\cite{nad} or the DCNQI materials \cite{hiraki}
(for a collection of recent results see
ref.\ \cite{brazovskii}).

We consider the Hamiltonian (\ref{eq:H-LR}) for large coupling constants,
$U\gg t, V\gg t$. Since the variational ansatz (\ref{eq:Psi_B}) does not
penalize configurations with doubly occupied sites, we use
a refined wave function \cite{Otsuka, Dzierzawa}
\begin{equation}
|\Psi_{BG}\rangle= e^{-\lambda \hat{D}}e^{-\eta \hat{T}}|\Psi_\infty\rangle
\;\;,
\label{psibg}
\end{equation}
where $|\Psi_\infty \rangle$ is now a linear
superposition of states that can be visualized in real space as
\begin{equation}
\Big\{ \quad|\quad \sigma_1
\quad 0 \quad \sigma_2 \quad 0
\quad \sigma_3 \quad 0 \quad \cdots
\quad \sigma_N \quad 0 \quad \rangle\Big\}\;\;.
\label{gwc}
\end{equation}
($\sigma_i$ is the spin of a localized
electron and zeroes stand for empty sites). Since
the classical ground state is highly ($2^N$) degenerate, we do not introduce
{\it a priori} any special magnetic order (e.g.\ antiferromagnetic or
ferromagnetic) but {\it we allow for all possible spin configurations}.
In the variational wave function
(\ref{psibg}) the term $e^{-\eta \hat{T}}$
again controls the delocalization of charge away from the
generalized Wigner lattice (\ref{gwc}) and
$e^{-\lambda \hat{D}}$ reduces the weight of configurations
with doubly occupied sites. As there is no prefered spin
configuration in the classical GWL, magnetic correlations will be generated
exclusively by quantum fluctuations.

\begin{figure}[htbp]
\resizebox{6cm}{!}{\includegraphics{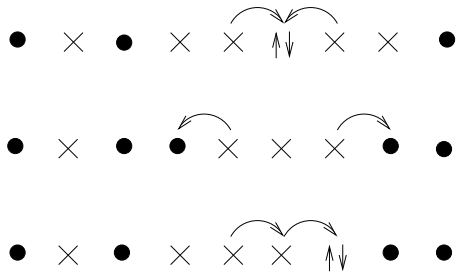}}
\caption{Intermediate state contributing 
to the interaction energy at fourth order. 
Dots/crosses stand for filled/empty sites respectively. 
 Notice that the first and third processes can give rise 
to spin exchange.}
\label{fig:procc3}
\end{figure}

The energy of the ground state is obtained by minimizing the expression
\begin{eqnarray}
\lefteqn{E_{BG}(\lambda,\eta)  =
\frac{\langle\Psi_{BG}|\hat{H}|\Psi_{BG}\rangle}
{\langle\Psi_{BG}|\Psi_{BG}\rangle}= {}}
\label{e-bg} \\
& & {} \frac{\langle\Psi_\infty|e^{-\eta \hat{T}}e^{-\lambda \hat{D}}
(-t\hat{T}+U\hat{D}+V\hat{W})e^{-\lambda \hat{D}}
e^{-\eta \hat{T}}
|\Psi_\infty\rangle}{\langle\Psi_\infty|e^{-\eta \hat{T}}
e^{-2\lambda \hat{D}}e^{-\eta \hat{T}}|\Psi_\infty\rangle}\;\; \nonumber
\label{mainresult}
\end{eqnarray}
with respect to the two variational parameters $\lambda$ and $\eta$.
Unlike what happens in the spinless case, a closed analytical expression
for $E_{BG}(\lambda,\eta)$ has not been found. Instead, we
use the expansion $e^{-\eta \hat{T}}\simeq 1 -\eta \hat{T}
+\frac{1}{2}\eta^2\hat{T}^2+\ldots$, which is valid for small $\eta$
(i.e.\ small $t$). \footnote{A similar expansion has been successfully used
for
the Hubbard chain at half filling, where one correctly obtains the Heisenberg
antiferromagnet with exchange constant $J=4t^2/U$. \cite{Baer}}
Each factor $\hat{T}$ produces hopping to neighboring sites. Therefore,
in order to allow for spin exchange, we have to expand at least to fourth order
in $\eta$.
The corresponding hopping processes are illustrated in figure \ref{fig:procc3}.
They lead to an exchange of neighboring spins as shown explicitly in
Appendix A, where the expectation values of the hopping term $\hat{T}$, of the
number of doubly occupied sites $\hat{D}$ and of the long--range part of the
interaction $\hat{W}$ are calculated to fourth order in $\eta$. Collecting
the results of eqs.\ (\ref{meankinetic}), (\ref{meandouble}) and (\ref{Vspin}),
we obtain the total energy
\begin{eqnarray}
\lefteqn{E_{BG}(\lambda,\eta)  =
L\big[-\frac{V}{4}\log 2 + 2\eta t+(2\log 2-1)\eta^2V\big] {}}
\nonumber \\
& & {}+J \sum_{i\ even}\langle\Psi_\infty|\vec{S}_i \cdot \vec{S}_{i+2}-
\frac{1}{4}n_i n_{i+2}|\Psi_\infty\rangle\ ,
\label{eq.mainresult}
\end{eqnarray}
where the exchange constant $J$ is given by
\begin{eqnarray}
J&=&-4\eta^3t(1+3e^{-\lambda})-3\eta^4Ue^{-2\lambda}\label{eq:exchange}\\
& & -4V\eta^4\big[2\log 2 -1 +e^{-2\lambda}\big(2\log 2-\frac{15}{8}\big) \big]\ .
\nonumber
\end{eqnarray}

We see that the variational parameters $\eta$ (which controls the
delocalization of charge) and $\lambda$ (which reduces double occupancy)
can be determined independently of the spin configurations contributing
to the magnetic ground state, which depends only on the sign of the
exchange constant. For antiferromagnetic exchange the ground state energy
is given by \cite{hulthen}
\begin{equation}
\sum_{i\ even}\langle\Psi_\infty|\vec{S}_i \cdot \vec{S}_{i+2}-
\frac{1}{4}n_i n_{i+2}|\Psi_\infty\rangle = -\frac{L}{2}\log 2\ .
\end{equation}
If the parameter $\eta$ is small enough (which is certainly the case in
the Wigner crystal phase), it is essentially determined by the minimum
of the first term in eq.\ (\ref{mainresult}), i.e.
\begin{equation}
\eta \approx -\frac{t}{(2\log 2-1)V}\ .
\end{equation}
On the other hand, $\lambda$ is chosen such as to maximize the exchange
constant $J$, namely
\begin{equation}
e^{-\lambda} = -\frac{3t}{2\eta[\frac{3}{4}U+(2\log 2-\frac{15}{8})V]}\ .
\end{equation}
Inserting these two values into eq.\ (\ref{eq:exchange}) we obtain
\begin{equation}
J\approx\frac{9t^4}{(2\log 2-1)^2V^2[\frac{3}{4}U+(2\log 2-\frac{15}{8})V]}\ .
\end{equation}
The resulting ground state energy per site
\begin{equation}
\frac{1}{L}E_{BG}=-\frac{V}{4}\log 2 -\frac{t^2}{(2\log 2 -1)V}
-\frac{J}{2}\log 2
\end{equation}
consists of three terms, the energy of the classical Wigner crystal,
the energy gain due to charge delocalization and the magnetic exchange
energy.

We see here that the finiteness of $U$ leads to an
antiferromagnetic coupling between the spins, which is a direct consequence
of the quantum fluctuations in the charge--ordered state.
Recently it has been shown that the
spin correlations in the Heisenberg chain decay like
$(-1)^{(i-j)} (\log|i-j|)^{1/2}/|i-j|$.\cite{auerbach}
In the framework of our variational approach,
the long--range order in the charge sector is therefore
accompanied by an algebraic magnetic order.
\footnote{Note that, in 
contrast to spin correlations, which vanish for $U\rightarrow \infty$,
density correlations depend only weakly on $U$ for small t (and
therefore small $\eta$). In fact, it is easy to show that up to second
order in $\eta$ the static structure factor is identical to that of
spinless fermions (i.e., to the limit $U\rightarrow \infty$).}

The variational analysis presented above for $n=1/2$ indicates that
in the region where the approach is valid, $U\gg t, V\gg t$, there is a
{\it separation of energy scales} between the charge and the spin
degrees of freedom. Recently, charge ordering has been observed
in the organic chain compounds (TMTTF)$_2$ PF$_6$ and
(TMTTF)$_2$ AsF$_6$ at $T_{CO}=70K$ and $100K$, respectively. No structural
transition has been observed so far at $T_{CO}$ and the magnetic
ordering occurs at much lower temperatures (around $10K$). It is tempting
to associate the charge ordering, which has been reported on the basis of both
$NMR$ \cite{Chow} and dielectric measurements,\cite{Monceau} with the
stabilization of a Wigner lattice, although for a detailed comparison with
experiments structural effects may have to be taken into account.
Previously, experiments on (DI-DCNQI)$_2$Ag have already been interpreted
in terms of a Wigner crystal.\cite{hiraki}

It has become popular to describe organic conductors in terms of the
extended Hubbard model, where only on--site and nearest--neighbor
Coulomb interactions with coupling constants $U$ and $V_1$, respectively,
are taken into account.\cite{Ung, mila95, Seo} Informations about the
parameter values can be obtained by exploiting the well--known sum rule
relating the optical absorption to the total kinetic energy. For
(TMTTF)$_2$PF$_6$ exact diagonalization studies for the kinetic energy
reproduce the observed reduction of oscillator strength of 0.73
\cite{Jacobsen}
(as compared to that of non--interacting electrons) using the parameters $U=8t$
and $V_1=3t$ or else $U=6.7t$ and $V=3.3t$
(see Fig.\ 4a in ref.\ \cite{mila95}). Our variational
approach applied to the same model would give roughly the same values
together with an exchange constant $J_1=6t^2/(UV_1^2)\approx 0.08t$.
For long--range Coulomb interactions, which suppress the kinetic energy
much less, the situation is quite different. In fact, for $n=1/2$ and
$U=2V$, a reduction factor of 0.73 would require a Coulomb interaction strength
of $V=9t$. Nevertheless, the exchange constant, given by
eq.\ (\ref{eq:exchange}) would again be of the order of $0.08t$.

In addition to the simple case presented here,
the study of more complex fillings can be envisaged.
For instance, the GWL configuration $110110\ldots$ at filling
$n\simeq 2/3$, which is
relevant to TTF-TCNQ, intrinsically leads to a dimerization of
the exchange constants, and could result in qualitatively new
physics. Our approach may also shed new light onto the $4k_F$ modulations
observed in TTF-TCNQ \cite{Pouget} and in several 1:2 compounds. Most of the
structural data exhibit a single harmonic modulation, so that they could be
attributed to region (III) of the ``phase diagram'' of
fig.\ \ref{fig:phasediag}.

\section{Conclusion}
\label{conclusion}

We have studied a one--dimensional system of electrons
interacting through the long--range Coulomb interaction on a discrete lattice.
Starting from the charge--ordered classical limit, we have
introduced a variational wave function which allows to treat the effects
of quantum fluctuations. This
wave function is always insulating and charge ordered  at any
commensurate filling,  and is
particularly well suited in the limit of strong interactions.

Although no phase transition arises within our treatment,
the phase diagram can be clearly divided into
three distinct regions separated by crossovers:
(I) the classical generalized Wigner lattice in the limit
where the band motion is negligible,
(II) a Wigner crystal regime at intermediate couplings and (III) a weakly
modulated regime at low interaction
strengths.  Lattice effects appear to be particularly important
at simple commensurate fillings, where
the charge fluctuations are reduced as compared to those of a continuous
system. This  general trend  agrees with recent results obtained in two space
dimensions,\cite{imada02} where it was shown that the critical parameter
for Wigner crystallization is dramatically
reduced from the value $r_s\approx 35$ in continuous models to
$r_s\approx 2$ in a quarter--filled band.

Concerning the spin degrees of freedom, exchange processes induced by
quantum fluctuations were shown to give rise to
antiferromagnetic correlations. These develop
out of the charge ordered configuration, with a much lower characteristic
energy
scale than typical Coulomb interaction energies.
For the quarter--filled band the spin sector has been
mapped onto the antiferromagnetic Heisenberg chain, which exhibits
algebraically decaying spin correlations.

The existence of a charge--ordered ground state, as well as the
aforementioned decoupling of magnetic/charge energy scales,
are not unique to long--range interactions. However, besides quantitative
differences between the genuinely long--ranged and the {\it ad hoc}
short--ranged models, taking into account the unscreened long--range
Coulomb potential appears to be crucial in the quest of
a unified view of the insulator--to--metal transition in
strongly correlated systems, since it is the only means to span all the
possible band fillings. In this spirit, the
Mott transition occurring at the ``most'' commensurate filling
$n=1$ and the Wigner crystal at $n\to 0$ are intimately connected,
and can be viewed as two limiting cases of the same general phenomenon.

\acknowledgments

We wish to thank F. Nad and J.-P. Pouget for fruitful discussions.
One of us (B.V.) is very grateful to M.A.H Vozmediano
for valuable discussions.
Financial support from MEC (Spain) through Grant No. PB96/0875 and
CAM (Madrid) through Grant No. 07/0045/98 is acknowledged.
This work has also been supported by the Swiss National Foundation
through grant no.\ 20-61470.00.

\appendix

\section{Variational energy at quarter--filling}
\label{app:spin}
In this appendix we will detail the computation of the
ground--state energy for the quarter--filled case $n=N/L=1/2$,
using the trial ground state (\ref{psibg}). We have to
evaluate the three terms
\begin{equation}
E_{BG}(\eta,\lambda)  =
-t\langle \hat{T} \rangle + U \langle \hat{D} \rangle +
V \langle \hat{W} \rangle\ ,
\label{app-ebg}
\end{equation}
where
\begin{equation}
\langle \hat{O} \rangle =
 \frac{\langle\Psi_\infty|e^{-\eta \hat{T}}e^{-\lambda \hat{D}}
\hat{O}e^{-\lambda \hat{D}} e^{-\eta \hat{T}}
|\Psi_\infty\rangle}{\langle\Psi_\infty|e^{-\eta \hat{T}}
e^{-2\lambda \hat{D}}e^{-\eta \hat{T}}|\Psi_\infty\rangle}\ .
\label{app-average}
\end{equation}
Besides the variational parameters $\eta$ and $\lambda$, the wave function
$|\Psi_\infty\rangle$ itself is not known {\it a priori}, but has also
to be determined by minimizing the energy. It will turn out that
$|\Psi_\infty\rangle$ is the ground state of the 1D Heisenberg antiferromagnet,
with spins located on even (or odd) sites. We use an expansion in powers of
$\eta$ and therefore have to assume that $V$ is large enough to keep the
electrons close to their positions in the classical limit. As the analysis
is straightforward, we only indicate the main steps. \cite{belen}

\subsection{Kinetic energy}
Expanding to fourth order in $\eta$ we obtain for the mean value of the
kinetic energy
\begin{eqnarray}
\langle \hat{T} \rangle & = & -2\eta
\langle\Psi_\infty|\hat{T}^2|\Psi_\infty\rangle-\frac{4}{3}\eta^3
\langle\Psi_\infty|\hat{T}^4|\Psi_\infty\rangle+{}
        \nonumber\\
 & & -{} \eta^3 \langle\Psi_\infty|\hat{T}^2(e^{-\lambda \hat{D}}-1)
\hat{T}^2|\Psi_\infty\rangle + {}
         \nonumber\\
& & {} + 4 \eta^3
(\langle\Psi_\infty|\hat{T}^2|\Psi_\infty\rangle)^2\ .
\label{meankin}
\end{eqnarray}
To proceed we have to compute correlation functions of the type
$\langle\Psi_\infty|\hat{T}^{2m}|\Psi_\infty\rangle$ (all contributions
with odd exponents vanish). The operator $\hat{T}^{2m}$ can
be decomposed using Wick's theorem.
The expectation value of the squared kinetic energy operator ($m=1$) is
easily computed and gives
$\langle\Psi_\infty|\hat{T}^2|\Psi_\infty\rangle=L$.
The expectation value of the fourth power of the kinetic energy
is more involved, but it is easy to see that only terms with
two or three contractions contribute. These can be expressed in terms
of density operators $n_i$ and spin $\frac{1}{2}$ operators ${\vec{S}}_i$,
giving
\begin{equation}
\frac{1}{3}\langle\Psi_\infty|\hat{T}^4|\Psi_\infty\rangle  =
L^2-\sum_{i\ even}\langle\Psi_\infty|4 \vec{S}_i \cdot \vec{S}_{i+2}-
n_i n_{i+2}|\Psi_\infty\rangle\ .
\label{t4}
\end{equation}
The term $L^2$ is cancelled by a corresponding term coming from the
denominator of eq.\ (\ref{app-average}), so that only ``linked diagrams''
contribute to the expectation value of the kinetic energy.
In order to calculate the remaining term in eq.\ (\ref{meankin}),
we notice that only the first and third processes in fig.\ \ref{fig:procc3}
contribute. We obtain
\begin{eqnarray}
\lefteqn{\left(e^{-\lambda \hat{D}}-1\right) \hat{T}^2|\Psi_\infty \rangle = }
\nonumber \\
& & (e^{-\lambda}-1)\Big[\sum_{i\ odd, s\sigma} c_{i\sigma}^+
c_{i+s\sigma}
c_{i\bar{\sigma}}^+c_{i-s\bar{\sigma}}|\Psi_\infty\rangle + \nonumber \\
& &  \sum_{i\ even, s\sigma} c_{i\sigma}^+
c_{i+s\sigma}
c_{i+s \sigma}^+c_{i+2s \sigma}|\Psi_\infty\rangle \Big] \ .
\label{cost-en}
\end{eqnarray}
Using again the Wick decomposition and the representation of the
remaining fermionic operators in terms of spin and density operators, we find
\begin{eqnarray}
\lefteqn{\langle\Psi_\infty|\hat{T}^2(e^{-\lambda \hat{D}}-1)
\hat{T}^2|\Psi_\infty\rangle =} \nonumber \\
& & -3(e^{-\lambda}-1)\sum_{i\ even}\langle\Psi_\infty|4 \vec{S}_i \cdot
\vec{S}_{i+2}-
n_i n_{i+2}|\Psi_\infty\rangle\ .
\label{eq:proc13}
\end{eqnarray}
Collecting all the terms in Eq.\ (\ref{meankin}) we arrive at
\begin{equation}
\langle\hat{T}\rangle=-2\eta L +\eta^3(1+3e^{-\lambda})
\sum_{i\ even}\langle\Psi_\infty|4 \vec{S}_i \cdot \vec{S}_{i+2}-
n_i n_{i+2}|\Psi_\infty\rangle\ .
\label{meankinetic}
\end{equation}

\subsection{On--site interaction energy}
The expansion in $\eta$ of the mean value of the on--site interaction energy
gives
\begin{equation}
\langle \hat{D} \rangle = \frac{1}{4} \langle \Psi_\infty | \hat{T}^2
\hat{D} e^{-2\lambda \hat{D}} \hat{T}^2 |\Psi_\infty \rangle \eta^4 +
O(\eta^6)
\end{equation}
and can be immediately evaluated by comparison with eq.\ (\ref{eq:proc13}),
\begin{equation}
\langle\hat{D}\rangle = -\frac{3}{4}\eta^4e^{-2\lambda}
\sum_{i\ even}\langle\Psi_\infty|4 \vec{S}_i \cdot \vec{S}_{i+2}-
n_i n_{i+2}|\Psi_\infty\rangle\ .
\label{meandouble}
\end{equation}

\subsection{Long--range potential energy}
To compute the expectation value $\langle \hat{W} \rangle$
of the long--range potential, we first recall that $|\Psi_{\infty}\rangle$
is an eigenstate of $\hat{W}$, namely (for an overall neutral system)
\begin{equation}
\hat{W}|\Psi_\infty\rangle= -\frac{L}{4}\log 2\ |\Psi_\infty\rangle\ .
\label{eq:wcl}
\end{equation}
For $\hat{W}'=\hat{W}+(L/4)\log 2$ all the terms
where $\hat{W'}$ acts directly on the wave function are thus suppressed
and we have to evaluate only four terms to order $\eta^4$,
\begin{eqnarray}
\lefteqn{\langle\Psi_\infty|e^{-\eta \hat{T}}
e^{-\lambda \hat{D}}
\hat{W}'e^{-\lambda \hat{D}} e^{-\eta \hat{T}}|\Psi_\infty\rangle
= \eta^2 \langle\Psi_\infty|\hat{T}\hat{W}'\hat{T}
|\Psi_\infty\rangle+{} }
         \nonumber\\
 & & {} +\eta^4
\Big\{\frac{1}{6}(\langle\Psi_\infty|\hat{T}\hat{W}'
\hat{T}^3|\Psi_\infty\rangle
+\langle\Psi_\infty|\hat{T}^3
\hat{W}'\hat{T}|\Psi_\infty\rangle)+{}
          \nonumber\\
  & & {} +\frac{1}{4}\langle\Psi_\infty|\hat{T}^2 e^{-\lambda \hat{D}}
\hat{W}' e^{-\lambda \hat{D}} \hat{T}^2 |\Psi_\infty\rangle\Big\}
+O(\eta^6)\;\;.
\label{devVbg}
\end{eqnarray}
In this case all the three processes in fig.\ \ref{fig:procc3} contribute.
An additional complication is the long--range nature of the potential which
requires the summation of infinite series. The first three terms of the
r.h.s.\
of eq.\ (\ref{devVbg}) can be linked to previously calculated expressions
in view
of the relation
\begin{equation}
\hat{W}'\hat{T}|\Psi_{\infty}\rangle = (2\log 2-1)\hat{T}|\Psi_{\infty}\rangle.
\end{equation}
The last term in eq.\ (\ref{devVbg}) is found to be
\begin{eqnarray}
\lefteqn{\langle\Psi_\infty|\hat{T}^2 e^{-\lambda \hat{D}}
\hat{W}' e^{-\lambda \hat{D}} \hat{T}^2 |\Psi_\infty\rangle =}
        \\
 & & {} 4[(2\log 2-1)L^2+(\frac{11}{2}-10\log 2)L] +
          \nonumber\\
  & & {}(\frac{15}{2}-8\log 2) e^{-2\lambda}
\sum_{i\ even}\langle\Psi_\infty|4 \vec{S}_i \cdot \vec{S}_{i+2}-
n_i n_{i+2}|\Psi_\infty\rangle\ .  \nonumber
\end{eqnarray}
Terms proportional to $L^2$ are again cancelled by terms from the denominator.
Adding the contribution (\ref{eq:wcl}) we finally get

\begin{eqnarray}
\lefteqn{\langle\hat{W}\rangle= -\frac{L}{4}\log 2 +L(2\log 2-1)\eta^2{}}
         \nonumber\\
 & & {} +\Big[1-2\log 2+\Big(\frac{15}{8}-2\log 2\Big)e^{-2\lambda}\Big]
\eta^4\nonumber\\
 & & {} \times\sum_{i\ even}\langle\Psi_\infty|4 \vec{S}_i \cdot \vec{S}_{i+2}-
n_i n_{i+2}|\Psi_\infty\rangle\ .
\label{Vspin}
\end{eqnarray}

\end{document}